# Recommender Systems for Software Project Managers

Liang Wei

Precima Inc., Toronto – Ontario – Canada, Leonard.wei96@gmail.com

Luiz Fernando Capretz

Department of Electrical & Computer Engineering – Western University, London – Ontario – Canada, lcapretz@uwo.ca

**Abstract**

The design of recommendation systems is based on complex information processing and big data interaction. This personalized view has evolved into a hot area in the past decade, where applications might have been proved to help for solving problem in the software development field. Therefore, with the evolvement of Recommendation System in Software Engineering (RSSE), the coordination of software projects with their stakeholders is improving. This experiment examines four open source recommender systems and implemented a customized recommender engine with two industrial-oriented packages: Lenskit and Mahout. Each of the main functions was examined and issues were identified during the experiment.

## 1 Introduction

Software engineering (SE) is a knowledge-oriented application that navigates intensive information in a systematic fashion. Information in SE includes the design document, requirements specifications, source code, task catalogue, meeting notes, work backlog, component technologies, learning resources, and human resources. The techniques, volumes, and dynamics of this information omits data that require a special application of classified intelligence to support developers. Robillard et al. [1] defines the Recommendation System in SE (RSSE) as a software application that provides potential items that are valuable for a SE task in the relevant context. The core challenge lies in how to establish context, which could include all relevant information about the user, related work environment, and the project status at the time of the recommendation. A tourist who seeks hotel recommendations can simply specify the context with a series of criteria, such as location, amenities, distance, and customer rating. In contrast, a project manager or technical lead who seeks suitable software aids for managing source code, documentation, communication channels, and task and defect tracking, needs a recommendation system that can establish fuzzy heuristics, such as which tools have been used in previous projects and which areas of business are related to the current needs.

The initiation process of a new project is highly important for all stakeholders; it contains outcomes that will be created and values that will be gained. Project manager (PM) selection is realized in the project initiation process and it is critical for project success [2]. The role of project manager could be critical to its success when a project is initiated, during which there is no historical data that can be of any help. This is referred to as a RSSE ramp-up problem – also called cold start problem – which denotes a situation where a RSSE needs initial data before the algorithms are stabilized to determine reasonable recommendations. The remainder of this paper is organized as follows: Section 2 summarizes branches in SE, introduces the cold start problem, and discusses challenges and ideas. Section 3 demonstrates four open source recommendation systems by benchmarking and comparing each of them and explain their implementation. Finally, Section 4 contains the conclusions of this analysis.

## 2 Recommender Systems in Software Engineering

Many recommendation systems have been brought to industry to help analysts meet the evolving, complex and heterogeneous needs of software systems in a variety of activities [3]. Many problems in SE are not constrained with data, but with the capability of computation. Considering the impact of change analysis [4], for the developer to determine the impact of changes is clear, but in general it is impossible to accurately calculate the solution. Thus, in the field of SE and in other technical fields, guidance in the form of recommendations is required not only to navigate large information spaces, but also to deal with forms of undecidable problems or problems in which exact solutions cannot be calculated in real time. A recommendation system supports the business by easily providing tasks or processes and uncertainties in decision making. All levels of organizations can take advantage of the output from recommended systems to overcome unknown challenges: managing resources, improving the quality of work, and exposing the code itself when solving problems.

## 2.1 Branches in Software Engineering

New project members are faced with a landscape of information about the project, which requires a period of time for them to get acquainted. Such information varies among organizations and project management approaches; however, the landscape will basically contain information from a number of sources.

- Project profile. Information about a project is stored and managed in the version control system (VCS) including code change, specification changes and ownership of the code or function unit. Unfortunately, information stored in a VCS is not designated for search and reuse purpose. Useful knowledge must often be inferred from the VCS and other repositories, typically by using a combination of heuristics and data mining techniques.
- Documentation archives. This provides formal deliverables and informal communication records (emails, change requests, etc.), to team members and other stakeholders. Higher-level business documents, such as a task distribution list and milestone planning aligned with assignees, give information about the types of techniques for a task the owner requires and the management tool being used.
- Usage feedback. Many applications intend to collect user behavioral data to improve the application interaction and user experience. User behavioral data is extracted from a completely unorganized log that records user behavior while using the application. In SE, these forms of data monitor development tendencies during the integration phase. Moreover, data collected during the execution phase contains a source of knowledge, particularly important for quality assurance purposes.
- Software APIs and related learning resources. A project team with experienced members always makes full use of reusable software assets (frameworks, codes, and function libraries) referenced to application programming interfaces (APIs). Unlike every line of code, APIs introduce an object-oriented, logically structured program architecture that requires developer comprehension of tasks. A well-maintained API and correlated applications are typically extended by informational documentation and a data dictionary, such as user manuals, and function descriptions.
- Development environment. The development environment includes all the configuration information, systematic structure, development tools, executing commands and test cases. Such a development environment can quickly become complex to the point where developers perform simply because they are unaware of the tools and commands at their disposal.

## 2.2 Cold Start Problem

Consider a developer who is in need of finding a source code (piece of code and functions, etc.) that is related to the present task. This is a nice-to-have feature that software developers frequently demand as they fix defects, enhance new features, or perform maintenance jobs. In fact, one study estimates that developers spend 60-90% of their time reading and navigating code [5]. However, finding functioning code is always time consuming and challenging. Several informal lexical searches are often insufficient since the search term may not be known as the keyword to the code entries. Even if the developer can find some of the relevant elements, research has shown that it is difficult for a developer to faithfully find all dependencies from an entity [6].

RSSE is task-centric, rather than user-centric. In reality, the user always knows the task better than the recommendation engine developer. This situation is reflected in the limited amount of personalization in the RSSE. It is an open question as to whether personalization is necessary, even in software engineering [7]. Nevertheless, the main idea of RSSE is to reduce the time and effort in problem solving and task completion.

## 2.3 Challenges and Ideas

The primary goal of the recommendation system in software engineering is to help developers make decisions by intelligently narrowing down all possible alternatives to solve specific problems and providing simple, easy-to-use and actionable advice. In doing so, the recommendation system should communicate with the existing system in order to collect data through a user-friendly interface, analyze patterns, run its forecasting engine, and store results and reports. These systems typically require calibration several times during their lifetime to update the input (data) or adjust the granularity of the output. However, in many software organizations, there are few people such as analysts who can understand the mechanism of recommendation systems, evaluate the output, make decisions on data updates and performance improvement, and make changes accordingly. Large organizations tend to have large independent repositories; large software depots provide rich input to the recommendation system engine, but can present different ideas when compensating with historical data to the system:

- In terms of project management, in the initial phase, a project manager may need RSSE to give a list of available tools for status tracking, team communication, file storage, and client communication and collaboration.
- One trend is the increasing use of social networks in big data. Github is a platform in which the user's personal characteristics can be used to navigate information. Therefore, a further convergence between the RSSE and the traditional recommendation system will be witnessed [8].
- A graph of relationships, is important. This is built by ranking the association rating via metadata information, e.g., when a project is being initiated, project managers need a set of different tools for task tracking, documentation archiving, file sharing, stakeholders communication, and customer relationship management (CRM). An interface of a dashboard where recommendations can be of paramount importance [9].
- Felfernig et al. [10] presented a vision of a "recommendation and decision-making support system" that will support individual and group activities and identify the dependencies between requirements. This is done by recommending a quality review of stakeholders, prioritizing requirements, and recommending current requirements for the task. Software practitioners can work on demand artifacts to maximize group agreements and identify sets of requirements for future releases. In other words, they envision a system that could help a wide range of tasks related to requirements engineering.
- Issue management in SE is similar to general task management, such as service organization. The issue report is similar to the baton in where different participants (e.g., developers, testers, and customers) take part in the task resolving contributions and the problem management system is the central node for assigning subtasks to the operator.

## 3 Experiment and Benchmarking

Recommendation systems in SE could be of assistance in the implementation of specific tasks. In many cases, helpful recommendations will be independent from the tasks involving a developer. The benefit of the system, from the user's perspective, is that they are convinced by factors derived from real data. With easy access to information and content by adjustment, providing a positive recommendation can affect the user's attitude to the application so that the relationship with the system is increased trust. The user tasks we recommend may have different goals. The target can vary from, for example, reducing the information overload, making search and explore to locate interesting projects easier, improving quality, and reducing the time of the decision-making process. Increased trust and confidence in the service could also be key factors [11].

### 3.1 Dataset

This experiment uses one of the data sets shared by a community-based open source project, FLOSSmole [12], which provides collaborative collection and analysis of free libraries and open source project data covering the period starting in 2004, and growing with more than 68,000 web-based operations every month. The data contains millions of open source projects, developers, communication logs, and implementation artifacts.

The data is collected from the Object Web (currently OW2) project which is an independent, global, open-source software community. The purpose of OW2 is to a) promote the open-source with generic industrial applications and cloud computing techniques, and b) foster an active and vibrant developer community. The OW2 development approach is flexible and component-based. Components include software models, protocols, and applications to integrate, service-driven platforms for industrial computing. The datamart of Object Web are collected by participants with comprehensive metadata of projects, developers who support this experiment by enough volume and relational data from both empirical and business aspect. There are general description of projects, profiles of developers, information about projects such as environment, status, operating system, topic, and intended audience. By using Lenskit and Mahout framework and API, this experiment explores the model of user-based collaborative filtering and item-based collaborative filtering algorithms [13].

## 3.2 Benchmark

This study benchmarks four commonly used open source recommendation systems, EasyRec, Lenskit, Mahout, Turi, from various aspects through hands-on experiment with OW2, is listed in Table 1.

EasyRec implemented recommendation engine that is leverage REST API to enable website deployment and interaction, see Fig. 1 below. After User Actions (viewing, purchasing or rating an item) are sent to the recommendation engine through the REST API, these actions will be stored in the database for specialists' periodical analysis in identifying patterns in order to generate recommendations. These recommended results can be easily accessed through web service API provided by EasyRec and presented to the user.

LensKit provides highly configurable interfaces and components; their implementation primarily consists of finishing LenskitConfiguration, building on Gradle build, and using a recommender. LensKit implements the recommendation engine by using the rating prediction composed of two tasks: the ItemScorer is responsible for computing some kind of score, and the RatingPredictor maps those scores onto the rating range. The item recommender returns a list, and most implementations order the list in decreasing order of relevance (recommendedness). Mahout provides a rich set of components where developers can implement a customized recommendation system from a set of pre-defined algorithms.

Mahout is dedicated to providing enterprise-wide machine learning components with performance, scalability, and flexible capability. A recommender is the core abstraction in Mahout – when given a data model, it can generate recommendations. The GenericUser-BasedRecommender or GenericItemBased-Recommender will be most likely used while constructing a recommender engine. Mahout also provides an evaluation package for the accuracy of recommend-ation based on user data. More advanced, Mahout is focused on implementing machine learning in big data as well, it provides a wide range of algorithms including classification, clustering, and recommendation.

**Table 1.** Recommendation systems benchmark

| Name | License | Language | Build Tool | Type |
|---|---|---|---|---|
| EasyRec | GPL v2 | Java | Web (Rest api) | Recommendation |
| Lenskit | LGPL v2.1 | Java | Gradle | Recommendation |
| Mahout | Apache 2.0 | Java | Maven | Machine learning |
| Turi | Apache 2.0 | Python | GraphLab | Machine learning |

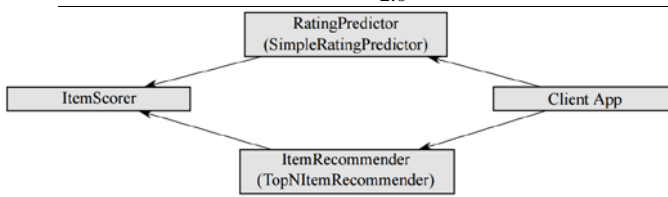

**Figure 1.** LensKit Recommender Structure

The Turi Machine Learning Platform provides various methods that enable data analysts and scientists to manipulate data in Tabular data and in Graphs, Visualization, and Featuring Engineering. Turi's GraphLab is a Python package that provides end-to-end data analysis and development. In the recommendation model, developers can choose working with either explicit or implicit data. With explicit data, a score for each interaction between a user and an item is given to an associated data column. Implicit data does not include rating, therefore, a dataset may only contain user ID and item ID. And the recommendations rely on the similarity of items and users that were interacting.

For deepened utilization and functional elaboration purposes, the matrix below (Table 2) shows a detailed function provided by each application specifically for building a recommender engine. As EasyRec allows third party web application to interact with data and recommendation on REST API, the data access is totally different from other three platforms. It empowers the recommender engine by enabling plugin features instead of built-in components, which are easy to deploy. In terms of algorithms, Lenskit and Mahout provide the same algorithms that are popularly used by recommendation systems. However Mahout is also able to handle big data inputs. Each application has limitations and minimum needs of input in order to get a desirably accurate prediction.

**Table 2. General Comparison of the Applications**

| Name | Limitation | Precision | Known problem |
| --- | --- | --- | --- |
| Turi | 5 or more records | -- | -- |
| Lenskit | -- | Requires 13 or more ratings per user for higher precision. | higher scores should be 'better' in whatever scorer measures |
| Mahout | Not all users get recommendations | -- | -- |

### 3.3 Implementation

The OW2 dataset contains relational data of 150 project details. From which, this experiment extracted data of five aspects: environment, program language, topic, operating system and intended audience. These five aspects become recommendation heuristics for offline recommender engine process, each relative count is regarded as ratings, e.g. developer Smith (dev_id = 3430) is a manager of project A.Smith (proj_id_num = 140) whose profile (see highlighted record in Figure 2) includes "audience = Developer, environment = Web Environment, operating system = OS Independent, Language = Java, Topic = Database Engines/Servers". Data in Figure 2 is processed into users who is taking project manager role since these users may contain more comprehensive heuristics for rating extraction.

In Lenskit, the recommender engine requires ratings and items files including userID, technologyID, preference count, and technology file, composed of technologyID, title, and genre. Lenskit provides basic item-item CF configuration that performs well on the widely used MovieLens data sets; it can be adapted to many different proposed variations and have components swapped out for new experiments. The ItemItemScorer class is the main driver for the item-item recommendation as the item scorer is decisive to that item-item CF algorithm configuration. The item-item CF model (similarity matrix) is built by ItemItemModelBuilder that specifies the similarity function used to compare items and inside it delegates to a general VectorSimilarity by default to compare the item's score (rating) vectors. Cosine or Pearson are used to allow both item and user similarities to reuse the same general vector similarity functions. Figure 2 is the recommendation for user 3430.

| dev_id | proj_id_num | audience | environment | OS_system | language | topic | PREFERENCE_COUNT |
|---|---|---|---|---|---|---|---|
| 14400 | 273 | Developers | Win32 (MS Windows) | OS Independent | Java | Code Generators | 5 |
| 10737 | 273 | Developers | Win32 (MS Windows) | OS Independent | Java | Code Generators | 5 |
| 14400 | 273 | Developers | Gnome | OS Independent | Java | Code Generators | 5 |
| 10737 | 273 | Developers | Gnome | OS Independent | Java | Code Generators | 5 |
| 14400 | 273 | Developers | KDE | OS Independent | Java | Code Generators | 5 |
| 10737 | 273 | Developers | KDE | OS Independent | Java | Code Generators | 5 |
| 3430 | 140 | Developers | Web Environment | OS Independent | Java | Database Engines/Servers | 5 |
| 13923 | 140 | Developers | Web Environment | OS Independent | Java | Database Engines/Servers | 5 |
| 3310 | 140 | Developers | Web Environment | OS Independent | Java | Database Engines/Servers | 5 |
| 7091 | 140 | Developers | Web Environment | OS Independent | Java | Database Engines/Servers | 5 |
| 9888 | 140 | Developers | Web Environment | OS Independent | Java | Database Engines/Servers | 5 |

**Figure 2.** Data of user '3430'

In Mahout, ratings file including userID, technologyID, and ratings is loaded to create the data Model in order for the recommender to handle interaction data. By creating a user-based recommender, system will compute recommendations for a particular user to look for others with a similar taste and pick the recommendations from their items. For finding similar users, the recommender engine has to compare their interactions. There are several methods among which this experiment uses Pearson Correlation Similarity to define user eligibility to leverage for the recommender. However, the dataset doesn't seem a good fit for user-based CF as the recommendations for most of users are not successful, 3 out of 103 users were generated, see Figure 3 for a segment of the recommendation output. While in applying the item-based recommender, it generated 100% recommendation for each user (Figure 4).

```
84 [main] DEBUG org.apache.mahout.cf.taste.impl.recommender.GenericUserBasedRecommender   - Recommendations are: [RecommendedItem[item:50087, value:5.0]]
RecommendedItem[item:50087, value:5.0]
84 [main] DEBUG org.apache.mahout.cf.taste.impl.recommender.GenericUserBasedRecommender   - Recommending items for user ID '291'
84 [main] DEBUG org.apache.mahout.cf.taste.impl.recommender.GenericUserBasedRecommender   - Recommending items for user ID '293'
85 [main] DEBUG org.apache.mahout.cf.taste.impl.recommender.GenericUserBasedRecommender   - Recommending items for user ID '303'
85 [main] DEBUG org.apache.mahout.cf.taste.impl.recommender.GenericUserBasedRecommender   - Recommendations are: []
85 [main] DEBUG org.apache.mahout.cf.taste.impl.recommender.GenericUserBasedRecommender   - Recommending items for user ID '312'
85 [main] DEBUG org.apache.mahout.cf.taste.impl.recommender.GenericUserBasedRecommender   - Recommending items for user ID '336'
85 [main] DEBUG org.apache.mahout.cf.taste.impl.recommender.GenericUserBasedRecommender   - Recommendations are: [RecommendedItem[item:20237, value:5.0]]
RecommendedItem[item:20237, value:5.0]
```

**Figure 3.** Recommendation in using user-based CF

```
130 [main] DEBUG org.apache.mahout.cf.taste.impl.recommender.GenericItemBasedRecommender   - Recommending items for user ID '2867'
130 [main] DEBUG org.apache.mahout.cf.taste.impl.recommender.GenericItemBasedRecommender   - Recommendations are: [RecommendedItem[item:20226, value:5.0]]
RecommendedItem[item:20226, value:5.0]
130 [main] DEBUG org.apache.mahout.cf.taste.impl.recommender.GenericItemBasedRecommender   - Recommending items for user ID '3310'
131 [main] DEBUG org.apache.mahout.cf.taste.impl.recommender.GenericItemBasedRecommender   - Recommendations are: [RecommendedItem[item:50066, value:5.0]]
RecommendedItem[item:50066, value:5.0]
131 [main] DEBUG org.apache.mahout.cf.taste.impl.recommender.GenericItemBasedRecommender   - Recommending items for user ID '3430'
131 [main] DEBUG org.apache.mahout.cf.taste.impl.recommender.GenericItemBasedRecommender   - Recommendations are: [RecommendedItem[item:50066, value:5.0]]
RecommendedItem[item:50066, value:5.0]
131 [main] DEBUG org.apache.mahout.cf.taste.impl.recommender.GenericItemBasedRecommender   - Recommending items for user ID '3751'
131 [main] DEBUG org.apache.mahout.cf.taste.impl.recommender.GenericItemBasedRecommender   - Recommendations are: [RecommendedItem[item:50066, value:5.0]]
RecommendedItem[item:50066, value:5.0]
131 [main] DEBUG org.apache.mahout.cf.taste.impl.recommender.GenericItemBasedRecommender   - Recommending items for user ID '3952'
132 [main] DEBUG org.apache.mahout.cf.taste.impl.recommender.GenericItemBasedRecommender   - Recommendations are: [RecommendedItem[item:50066, value:5.0]]
RecommendedItem[item:50066, value:5.0]
```

**Figure 4.** Recommendation using item-based CF

## 4 Conclusion

This paper mainly presents classified software products based upon their characteristics and uses them to propose a rule-based recommendation system. The recommendations provided can be useful to software developers in selecting the most appropriate software development life cycle model to use for the development of a software product. Primarily, the main contribution of this paper is the benchmark metrics and comparison through experimental practice for academic starters to have a comprehensive understanding of the resources they have before starting their own research, we believe it defines several to-be solved problems. In summary, this is a practical study that integrates open source datasets, codes, libraries, and tools like Microsoft SSIS.

Among the four major machine learning techniques discussed in the paper, KNN classifier algorithm is widely used by the four open source platforms in implementing recommender engine. Slope-one function can be found in EasyRec, Lenskit, and Mahout, meaning it is also a popular alternative for calculating recommendations. EasyRec is the only platform providing direct user access through REST API and data connection via JSON, which supports most of the website applications. In implementing Mahout, most of the effort will be given to converting the item-rating file into complied format for building a data model, which is a prerequisite for calculating the similarity and classification. In Lenskit, developers will have to perform data processing on item genre file and specific configuration file to guide the recommender engine. On the contrary, recommendation results of Mahout may not be stabilized and successful for each user. Lenskit can produce results on each user with given ratings and genre files along with a customized configuration file. This proves the findings in Table 1, that Mahout is a general machine learning platform containing recommendation methods, while Lenskit specializes in recommendations.

For software that is seeking middleware for an application recommendation engine, Lenskit may be the recommended choice if one only needs an add-on feature for recommendation. EasyRec would also be a wise option for a website that provides REST API to better integrate the recommendation feature with user experience. For medium or large organizations who may extend their data analysis area in big data, Mahout would obviously be the best choice for the functionality it may enhance and the ability of distributed computing it empowers.

This experiment was mainly developed with Java using Eclipse, MySQL, and Microsoft SSIS. OW2 data was imported into MySQL database; the original data was processed and extracted manually to generate the user-item rating file and genre. Microsoft SSIS was utilized to enhance the data transformation, which might not be easy to implement in MySQL.

The SE environment encounters significantly different contexts from the non-technical areas. Context-sensitive, dynamicity, and heterogeneity are the three main challenges of analyzing, interpreting, and evaluating the quality of SE projects. The limitation of data computation faced with the SE will add complexity to the machine learning techniques. These are the challenges faced by those who want to use software to organize their data and maneuver their business. A Recommendation System in Software Engineering is inevitably becoming one of the keys to successfully coping with these challenges.

**Table 3.** Technical and algorithm comparison

| Name | Algorithm | Model | Data Access |
| --- | --- | --- | --- |
| EasyRec | Mining association rules<br>Slope-One (plugin)<br>User Profile Aggregator (plugin) | Ranking<br>Clustering<br>(plugin interface enabled) | JSON<br>XML |
| Lenskit | kNN Item-based CF<br>kNN User-based CF<br>Matrix factorization<br>Slope-One | Scoring Items<br>Predicting Ratings<br>Top-N Recommendation | CSV Files (delimited text)<br>Databases via JDBC |
| Mahout | Item-based CF<br>User-based CF<br>Slope-One<br>SVD | Spark-item similarity<br>Neighborhood<br>Correlation<br>Preference Inferer | CSV Files<br>Databases via JDBC<br>Distributed (MapReduce) |
| Turi | Factorization machine<br>Matrix factorization<br>Item similarity<br>Popularity-based<br>Explicit and implicit | Rating acceptance<br>Item similarity<br>Model evaluation<br>Interactive visualizations | CSV Files<br>JSON<br>Databases via ODBC<br>Distributed (Spark RDD) |

*Proceedings of the Evaluation and Assessment in Software Engineering (EASE),* Trondheim, Norway, pp. 412-417, DOI: https://doi.org/10.1145/3463274.3463951, June 2021.# References

[1] M. P. Robillard, W. Maalej, R. J. Walkers, T. Zimmermann, "Recommendation Systems in Software Engineering," Springer Berlin Heidelberg, doi: 10.1007/978-3-642-45135-5, 2014.

[2] M., Radu, and R. Nistor, "Project initiation and project management approach – an expensive correction," Managerial Challenges of the Contemporary Society Proceedings, Vol. 6, 2013, pp. 67-70.

[3] M.P. Robillard, R.J. Walker, T. Zimmermann, "Recommendation systems for software engineering," IEEE Software, 27(4). doi:10.1109/MS.2009.161, 2010, pp. 80-86.

[4] A. Begel, R. DeLine, "Codebook : Social Networking over Code," Software Engineering - Companion Volume, 2009. ICSE-Companion of 31st International Conference on Software Engineering, May 2009, pp. 263-266.

[5] T. Zimmermann, P. Weißgerber, "Preprocessing CVS data for fine-grained analysis," Proceedings of the International Workshop on Mining Software Repositories, pp. 2-6, 2004.

[6] H. Reid, W.Robert, "Systematizing pragmatic software reuse," ACM Trans. Software Eng. Meth. 21(4), doi: 10.1145/2377656.2377657, 2012, vol. 20, pp. 1–44.

[7] B. Dagenais, H. Ossher, R. K. E. Bellamy, M. P. Robillard, and J. P. De Vries, "Moving into a new software project landscape," Proceedings of the ACM/IEEE 32nd International Conference on Software Engineering, 2010, vol. 1, pp. 275–284.

[8] E. Murphy-Hill, R. Jiresal, G. C. Murphy, "Improving software developers' fluency by recommending development environment commands," Proceedings of the ACM SIGSOFT International Symposium on Foundations of Software Engineering, doi: 10.1145/2393596.2393645, 2012, pp. 42:1-11.

[9] S. L. Lim, D. Quercia, and A. Finkelstein, "StakeNet: using social networks to analyse the stakeholders of large-scale software projects," Proceedings of the ACM/IEEE 32nd International Conference on Software Engineering, 2010, vol. 1, pp. 295–304.

[10] A. Felfernig, G. Friedrich, D. Jannach, M. Zanker, "An integrated environment for the development of knowledge-based recommender applications," International Journal on Electronic Commerce 11(2), 2006, pp. 11–34.

[11] R. Hu, "Design and user issues in personality-based recommender systems," Proceedings of the ACM Conference on Recommender Systems, Barcelona, Spain, doi:10.1145/1864708.1864790, September 2010, pp. 357-360

[12] J. Howison, M. Conklin, K. Crowston, "FLOSSmole: A collaborative repository for FLOSS research data and analyses," International Journal of Information Technology and Web Engineering, 2006, Vol. 1(3), pp. 17–26.

[13] E. Duala-Ekoko and M. P. Robillard, "Asking and answering questions about unfamiliar APIs: An exploratory study," Proceedings of the 34th ACM/IEEE International Conference on Software Engineering, 2012, pp. 266–276.